\documentclass[preprint,review,12pt]{elsarticle}

\usepackage[T1]{fontenc}
\usepackage{lmodern}
\usepackage{pbox}
\usepackage{array}
\usepackage{makecell}
\usepackage{multirow}
\usepackage{dsfont}
\usepackage{url}
\usepackage{subfig}
\usepackage{amsmath,amsfonts,amssymb, amsthm,bm}
\usepackage{graphicx}
\usepackage{float}
\usepackage{booktabs,tabularx}

\usepackage{stfloats}
\usepackage[toc,page]{appendix}

\usepackage{subfloat}

\usepackage{pbox}
\usepackage{array}

\begin{document}

\begin{frontmatter}

\title{Modelling Bitcoin Returns Volatility using GARCH models}

 \author[label1]
 {Samuel Asante Gyamerah}
 \address[label1]{Pan African University, Institute for Basic Sciences, Technology, and Innovation, Kenya}
 \tnotetext[label1]{correspondence:\,\, saasgyam@gmail.com}

\begin{abstract}
In this paper, an application of three GARCH-type models (sGARCH, iGARCH, and tGARCH) with Student t-distribution, Generalized Error distribution (GED), and Normal Inverse Gaussian (NIG) distribution are examined. The new development allows for the modeling of volatility clustering effects, the leptokurtic and the skewed distributions in the return series of Bitcoin. Comparative to the two distributions, the normal inverse Gaussian distribution captured adequately the fat tails and skewness in all the GARCH type models. The tGARCH model was the best model as it described the asymmetric occurrence of shocks in the Bitcoin market. That is, the response of investors to the same amount of good and bad news are distinct. From the empirical results, it can be concluded that tGARCH-NIG was the best model to estimate the volatility in the return series of Bitcoin. Generally, it would be optimal to use the NIG distribution in GARCH type models since time series of most cryptocurrency are leptokurtic. 
\end{abstract}


\end{frontmatter}


\section{Introduction}
\noindent
 The price of Bitcoin are known to be very volatile. Market players and investors are however interested in estimating accurately the volatility of Bitcoin. This is as a result of the correlation between volatility and returns on investment. It is notable that volatility are not directly observable and consequently the need for efficient model that can capture the volatility in the prices of Bitcoin. Estimating the volatility of Bitcoin is of high importance since Bitcoin has the highest market capitalization in the cryptocurrency market1. The cryptocurrency market is relatively new, hence there have not been enough literatures on the different models used in estimating the volatility in the market.
\\
[2mm]
\cite{gronwald2014economics} modelled Bitcoin price data using an autoregressive jump-intensity GARCH model and a standard GARCH model. Their results indicated that the autoregressive jump-intensity GARCH model performed better in fitting the Bitcoin price data than the standard GARCH model. \cite{gronwald2014economics} used multiple threshold-GARCH and Asymmetric-power GARCH to measure and estimate the volatility in the price of Bitcoin. They used Akaike information criterion (AIC), Bayesian information criterion (BIC) and Hannan-Quinn information criterion (HQC) to capture the leverage and regime switching features of the conditional variance and concluded that Bitcoin market is not yet mature. Using the price of Bitcoin, \cite{katsiampa2017volatility} studied the best conditional heteroscedasticity model in relation to goodness-of-fit. They revealed that the optimal model was the the AR-cGARCH model. \cite{chu2017garch} fitted 12 GARCH-type models to the log returns of the exchange rates of seven different cryptocurrencies including Bitcoin. IGARCH (1, 1) with normal innovations was found to be the optimal model using information criterion. Their conclusion however differs from the known stylized fact{\footnote{returns have fatty tails and high kurtosis}} of financial time series data.   
\\
[2mm]
Most financial time series data have shown that the conditional distribution of returns series exhibit several stylized features such as excess kurtosis, negative skewness, and temporal persistence in conditional movements. These stylized facts and properties of the returns seris have significant ramification for financial models especially volatility prediction, risk-scenario distribution, and in the case of financial crisis. To capture these stylized facts in these datas, econometricians have developed different tools to model and estimate volatility. 
\\
This paper study the estimation capacity of different nonparametric GARCH-type models on volatility of the return series of Bitcoin for the period between 01 January, 2014 to 16 August, 2019. Nonparametric methods are applied to the GARCH-type models because they do not assume any distributional assumptions and can capture the kurtosis and fatty tails of the return series of Bitcoin. Hence, we employ the Students-t distribution, Generalized Error Distribution, and the Normal Inverse Gaussian distribution.

\section{Materials and Methodology}
\subsection{Data}
\noindent
In this study the closing price is selected as the price of a cryptocurrency because it reflects all the activities of the cryptocurrency of the day. Historical daily closing price of Bitcoin was extracted from \url{https://coinmarketcap.com/}. The data extracted spanned from 01/01/2014 to 16/08/2019, totaling 2054 trading days. Bitcoin is a consensus network that enables a new payment system and a completely digital currency. 
\\
Assume $P_t$ and $P_{t-1}$ represents the current day and previous day price of a cryptocurrency, then the return series/log returms $(R_t)$ is calculated as 
\begin{equation}
R_t = log(P_t) - log(P_{t-1})
\label{log_returns}
\end{equation}

\subsection{Test of Normality}
\noindent
 The residuals data of the three cryptocurrency are checked for normality.  If the dataset is normally distributed, then a parametric statistic like the Normal distribution can be assumed. However, if the residuals data are not normally distributed, a non-parametric statistic will be used. If the normality test fails, it is important to consider the histogram and the normal probability plot to check the presence of outliers in the data set. We employed the Jacque-Bera (JB) test statistics and the Anderson-Darling (AD) test statistics for the normality test. The JB and the AD test statistics are defined as in equation \ref{JB_test} and \ref{AD_test} respectively: 
\begin{equation}
JB = n\Big(\dfrac{S^2}{6} + \dfrac{(K -3)^2}{24}\Big),
\label{JB_test}
\end{equation}
where $n$ is the sample size, $S$ and $K$ are the sample skewness and kurtosis respectively.
\begin{equation}
AD = -n -\dfrac{1}{n} \sum_{i=1}^{n} (2i - 1) [In F(Y_i) + In(1-F(Y_{n-i+1}))],
\label{AD_test}
\end{equation}
where $F(\cdot)$ is the cumulative distribution function (CDF) for the specified distribution and $i$ is the $i_{th}$ sample. 
\\ 
The JB test \cite{jarque1987test} uses the Lagrange multiplier approach on the Pearson family of distributions to derive tests for normality if observations and regression residuals. It is centered on sample skewness and kurtosis. The AD test \cite{anderson1954test} was developed by Anderson and Darling in 1954. It gives more weight to the tails of distribution. AD test is based on the empirical distribution function. The JB and AD test compare the scores in the sample to a normally distributed set of scores with same standard deviation and mean.  The null hypothesis of these test is "sample distribution is normally distributed". The distribution is not normally distributed if the test is significant.

\subsection{Testing for auto regressive conditional heteroscedasticity (ARCH) Effects}
\noindent
To apply GARCH models to the Bitcoin returns series, we test for the presence of stationarity and ARCH effects in the residual return series. The Ljung-box and Lagrange multiplier (LM) test \cite{engle2001garch} are used to test for the presence of ARCH effects in the data.

\subsubsection{Ljung Box test}
\noindent
The Ljung Box test \cite{box1970distribution} is a method used in testing the absence of serial autocorrelation up to a specified lag $k$. It can be defined as 
\\
$H_0$: The observed data do not exhibit serial correlation\\
$H_1$: The observed data exhibit serial correlation
\\
The test statistics is given as:
\begin{equation}
Q = N(N+2) \sum_{k=1}^{h} \dfrac{\hat{\rho}_k^2}{N-k}
\label{ljung}
\end{equation}
where $N$ is the sample size, $h$ is the time lag, $\hat{\rho}_k^2$ is the estimated autocorrelation of the series at lag $k$. 

\subsubsection{Lagrange multiplier (LM) test}
\noindent
To apply GARCH models to the returns series data, it is important to test the residuals for the presence of AutoRegressive Conditional Heteroscedasticity (ARCH) effects. The LM test was used to test for the presence of ARCH effects. LM is defined as;
\\
$H_0$: No ARCH effect\\
$H_1$: Presence of ARCH effects.
\\
The LM test statistics is given as:
\begin{equation}
LM = N.R^2 \sim \chi^2(q),
\label{LM_test_stats}
\end{equation}
where $N$ is the total number of observations, $R^2$ forms the regression, and $q$ is the number of restrictions. 

\subsection{Conditional variance equation}

\subsubsection{Standard Generalized AutoRegressive Conditional Heteroskedasticity (sGARCH)}
\noindent
GARCH is an extension of the ARCH model that integrates a moving average (MA) part together with the autoregressive (AR) part.
\\
Define a GARCH (p,q) model as 
\begin{equation}
\begin{aligned}
x_t &= \sigma_t \epsilon_t
\\
\sigma_t^2 &= \omega + \sum_{i=1}^{p} \alpha_i x_{t-i}^2  + \sum_{j=1}^{q} \beta_j \sigma_{t-j}^2
\end{aligned}
\label{GARCH}
\end{equation}
where $\omega > 0$, $\alpha_i > 0$, $\beta_j > 0$, $\sum_{i=1}^{p} \alpha_i + \sum_{j=1}^{q} \beta_j <1$. $\epsilon_t$ is the is a sequence of independent and identically distributed random variables. For the sake of tractability, the order of all the GARCH models used will be restricted to one.  
\\
The standard GARCH model \cite{bollerslev1986generalized} represented as sGARCH(1,1) is given as
\begin{equation*}
\sigma_t^2 = \omega+ \alpha_1 \epsilon_{t-1}^2 + \beta_1 \sigma_{t-1}^2
\end{equation*}

\subsubsection{Integrated GARCH (iGARCH)}
\noindent
Suppose the sum of the coefficients in equation \ref{GARCH} is equal to 1, the GARCH process is called an Integrated GARCH (iGARCH). That is, the iGARCH models are unit-root GARCH models. A basic characteristic of the iGARCH model is that the effect of past squared shocks $\eta_{t-i} = x_{t-1}^2 - \sigma_{t-i}^2$ for $i > 0$ on $x_t^2$ is persistent. An iGARCH(p,q) model assumes the form 
\begin{equation}
\begin{aligned}
x_t &= \sigma_t \epsilon_t
\\
\sigma_t^2 &=  \omega + \sum_{i=1}^{p} \beta_i \sigma_{t-i}^2 + \sum_{j=1}^{q}(1-\beta_j)\alpha_{t-j}^2
\end{aligned}
\label{IGARCH}
\end{equation}
Define an iGARCH (1,1) model as 
\begin{equation}
\begin{aligned}
x_t &= \sigma_t \epsilon_t
\\
\sigma_t^2 &= \omega + \beta_1 \sigma_{t-1}^2 + (1 - \beta_1) x_{t-1}^2, 
\end{aligned}
\label{igarch}
\end{equation}
where $1 > \beta_1 > 0$. 

\subsubsection{Threshold GARCH (tGARCH)}
\noindent
The tGARCH model was developed independently by \cite{ravichandran1989threshold} and \cite{glosten1993relation}. A tGARCH(p,q) model assumes the form
\begin{equation}
\sigma_t^2 = \ + \sum_{i=1}^{q}(\alpha_i + \lambda_i \mathds{1}_{t-i})x_{t-i}^2 + \sum_{j=1}^{p} \beta_j \sigma_{t-j}^2
\label{tgarch_form}
\end{equation}
where 
\begin{equation*}
\mathds{1}_{t-1} =
\begin{cases}
 1, \qquad \mbox{if} \quad  \epsilon_{t-i} < 0,  \qquad \mbox{bad news}
 \\
 0,  \qquad\,\, \mbox{if}\quad \epsilon_{t-i} \ge 0, \qquad \mbox{good news}
\end{cases}
\end{equation*}
$\alpha_i >0$, $\beta_j > 0$, and $\lambda_i > 0$.
\\
The generalized version of the tGARCH TGARCH(1,1) is given as:
\begin{equation}
\sigma_t^2 = \omega + \alpha x_{t-1}^2 + \beta \sigma_{t-1}^2 + \lambda x_{t-1}^2 \mathds{1}_{t-1}
\label{tgarch}
\end{equation}
where $\alpha$ and $\alpha + \lambda$ denote the effect of good news and bad news respectively. A $\lambda>0$ is an evidence that bad news upsurge volatility in the Bitcoin market. This indicates the existence of leverage effects of the first order. For an asymmetric news effect, $\lambda \neq 0$.

\subsection{Distribution Assumptions of the Error ($\epsilon_t$)}
\noindent
From the test of normality for the returns residuals in table \ref{JB_AD_test_statistics} and figure \ref{returns_residuals_plot}, it is clear that the distribution of the residuals returns are not normally distributed. There is the presence of excess kurtosis and heavy-tails in the distribution of the residuals returns. To account for the excess kurtosis and fat tails that are present in the residuals of the returns series, we model the error term in the GARCH models with a Student-t distribution, Generalized Error Distribution (GED), and a Normal Inverse Gaussian (NIG) types of distributions. These distributions are appropriate to capture the excess kurtosis and the skewness in the residuals return series.

\subsection{Student's t-distribution}
\noindent
The Student's t-distribution or the t-distribution is a sub-class of the continuous probability distributions. It is used when the sample size is small and the population standard deviation is unknown. It was proposed by \cite{student1908probable}. The probability density function (pdf) of the t-distribution is defined as 
\begin{equation}
f_t(x;v) = \dfrac{\Gamma(\frac{v+1}{2})}{\sqrt{v \pi}\Gamma(\frac{v}{2})}\Big(1 + \dfrac{x^2}{v}
\Big)^{-\frac{v+1}{2}}, \qquad x \in (-\infty, \infty)
\label{t_dist}
\end{equation}
where $v$ is the number of degrees of freedom and $\Gamma(\cdot)$ is the gamma function.
\\
The log likelihood function is defined as below
\begin{equation}
l(x;v) = N\,\mbox{In}\,\Bigg(\dfrac{\Gamma(\frac{v+1}{2})}{\sqrt{v \pi}\Gamma(\frac{v}{2})}\Bigg) - \dfrac{1}{2} \sum_{t=1}^{N} \Bigg( \,\mbox{In}\, \sigma_t^2 + (v+1) \,\mbox{In}\,\Bigg(1 + \dfrac{x_t^2}{\sigma_t^2(v-2)}\Bigg)\Bigg]
\label{log_t_dist}
\end{equation}

\subsubsection{Generalized Error Distribution (GED)}
\noindent
The standardized GED or the error distribution is a symmetrical unimodal sub-class of the exponential family with shape parameter $v$. The pdf of the GED is defined as in equation \ref{ged}, 
\begin{equation}
f_{GED}(x; v) = k(v) \exp \left\{  -\dfrac{1}{2} \Big| \dfrac{x}{\lambda(v)} \Big|^v
\right\}, \qquad x \in (-\infty, \infty)
\label{ged}
\end{equation}
where $v$ determines the tail weight with larger value of $v$ giving lesser tai weight, $k(v)$, and $\lambda(v)$ are constant and are defined as
\begin{equation*}
\begin{aligned}
k(v) &= \dfrac{v}{\lambda(v)2^{1+\frac{1}{v}} \Gamma\Big(\dfrac{1}{v}\Big)}
\\
\lambda(v) &=  \left\{ \dfrac{2^{-\frac{2}{v}\Gamma(\frac{1}{v})}}{\Gamma(\frac{3}{v})}
\right\}^{1/2}
\end{aligned}
\end{equation*}
$\Gamma(x) = \int_{0}^{\infty} t^{x-1}e^{-t}dt, \,\, x>0$ is the Euler gamma function.
\\
The log likelihood function is defined as 
\begin{equation}
l(x;v) = \sum_{t=1}^{N} \Bigg[ \mbox{In} \Big(\dfrac{v}{\lambda}\Big) - \dfrac{1}{2}\Big| \dfrac{x}{\sigma_t^2 v} \big|^v - (1+v^{-1})\,\mbox{In}\,2- \mbox{In}\,\Gamma \Big(\dfrac{1}{v}\Big) - \dfrac{1}{2}\,\mbox{In}\, \sigma_t^2 \Bigg]
\label{ged_log}
\end{equation}

\subsubsection{Normal Inverse Gaussian (NIG) Distribution}
\noindent
The NIG distribution was proposed by \cite{barndorff1977exponentially} and it is a sub-class of the generalized hyperbolic distribution. The pdf of an NIG distribution for a random variable $X$ is defined as
	\begin{equation}
	f_{NIG}(x; \alpha, \kappa,  \mu, \delta)) = \dfrac{\alpha \delta}{\pi} \exp \{\delta \sqrt{\alpha^2 - \beta^2} + \beta(x - \mu)\} K_1[\alpha q(x)]
	\label{NIG}
	\end{equation}
where $q(x) = ((x - \mu)^2 + \delta^2)^{1/2}$, $\alpha >0$ determines the shape, $0 \le \mid \kappa \mid \le \alpha$ determines the skewness, $\delta>0$ is the scaling, $\mu \in \mathbb{R}$ is the location parameter and $x \in \mathbb{R}$. $k_1(x)$ is the modified Bessel function of the second kind with index 1. The NIG distribution has a heavier tail than the normal distribution and can take different kinds of shapes. From equation \ref{NIG}, the shape of an NIG density is described using a four dimensional parameters ($\alpha, \kappa,  \mu, \delta$). The NIG distribution is appropriate for capturing data sets with extremal observations, skewness, and fat tails or semi-heavy tails. The log-likelihood of the NIG distribution is given as 
\begin{equation}
l (x; \alpha, \kappa, \mu, \delta) =  n \log \Big(\dfrac{\alpha \delta}{\pi}\Big) + n \delta \sqrt{\alpha^2 - \kappa^2} + \kappa \sum_{t=1}^{N} (x_t - \mu)  +  \sum_{t=1}^{N} \log K_1(x_t; \alpha, \delta, \mu)
\label{nig_likelihood}
\end{equation}

%

\subsection{Model selection}
\noindent
The best model was selected using two information criteria: the Akaike information criteria (AIC),
Bayesian information criteria (BIC). AIC and BIC considers the accuracy of the model fit and the number of parameters in the model; rewarding a better fit and penalizing an increased number of parameters in the return series data. The optimal model that will be selected is the model with the minimum AIC and BIC values.  The GARCH models used in this study were fitted using the maximum likelihood method. 

\begin{equation*}
AIC = -2\,\,\mbox{In}\,\,\mbox{L} (\hat{\Theta}) + 2k,
\label{aic}
\end{equation*}
\begin{equation*}
BIC = -2\,\,\mbox{In}\,\,\mbox{L}  + k\,\,\mbox{In}\,\,\mbox{L} (\hat{\Theta}),
\label{bic}
\end{equation*}
where $n, k, n, \Theta, \hat{\Theta}$ represent the number of observations, number of unknown parameters, vector of the unknown parameters, and the maximum likelihood estimates of the  vector of the unknown parameters respectively.\\
The smaller the AIC and BIC values of a model, the better the fit of that model.

\section{Results and Discussion}
\subsection{Descriptive Statistics}
\begin{figure}
	\centering
	\includegraphics[height=11cm,width=13cm]{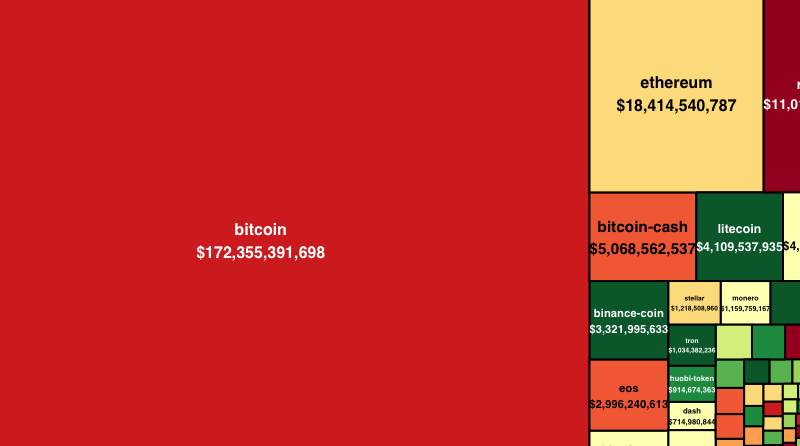}
	\caption{Cryptocurrencies by market capitalization}	
	\label{market_cap}
\end{figure}

\begin{figure}[H]
	\centering
	\includegraphics[height=07cm,width=14cm]{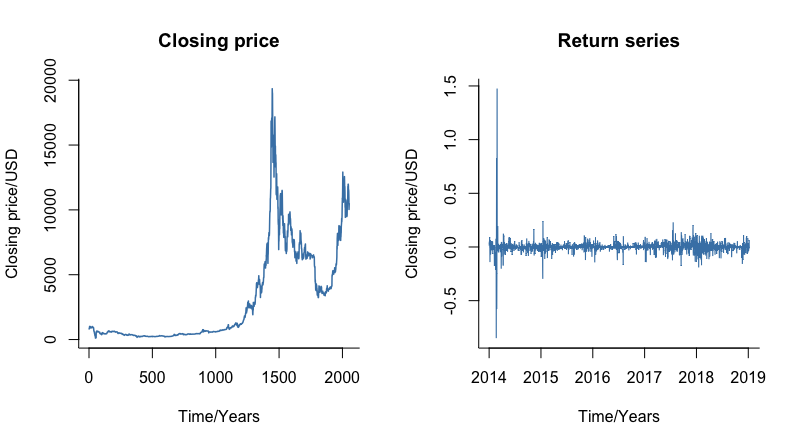}
	\caption{Closing price and return series of Bitcoin}	
	\label{close_return}
\end{figure}

\begin{figure}[H]
	\centering
	\includegraphics[height=08cm,width=14cm]{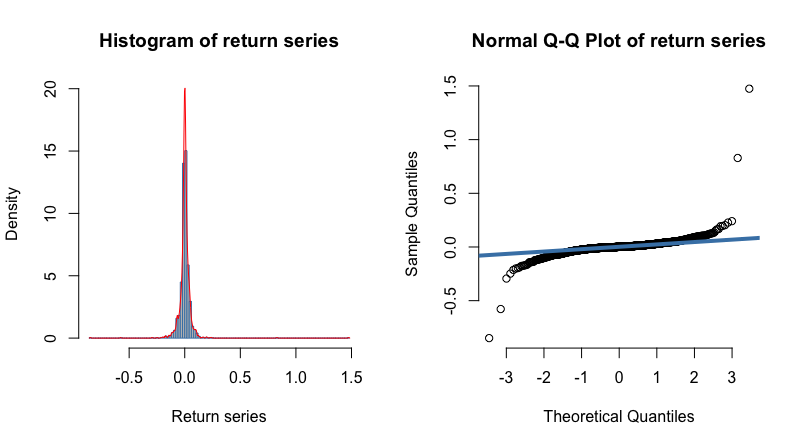}
	\caption{Histogram and normal q-q plot of return series of Bitcoin}	
	\label{hist_normal_returns}
\end{figure}

\begin{figure}[H]
	\centering
	\includegraphics[height=08cm,width=14cm]{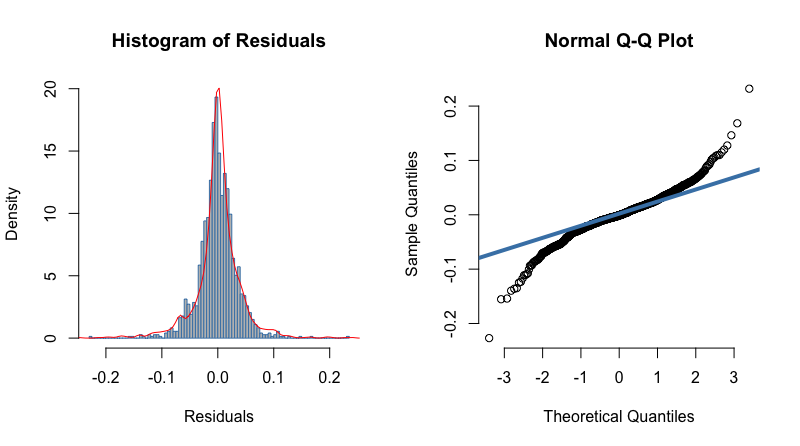}
	\caption{Histogram and normal q-q plot of return series of Bitcoin}	
	\label{returns_residuals_plot}
\end{figure}

\noindent
The visualization in figure \ref{market_cap} revealed that the entire cryptocurrency market is propped primarily by Bitcoin. Table \ref{bitcoin_info} is the statistics of Bitcoin on the cryptocurrency market and table \ref{descriptive_statistics} presents the descriptive statistics of the return series of the three cryptocurrencies. The return series of Bitcoin is positively skewed. These indicate that the returns of Bitcoin is non-symmetric. The positive value of the skewness indicates that the distribution of Bitcoin return series is skewed to the right or positively skewed. The positive excess kurtosis ($216.7461$) of the Bitcoin returns series indicates that the returns are leptokurtic. That is, the returns series has a fatty tail. Figure \ref{close_return} shows the time series plot of Bitcoin price (left figure) and the return series (right figure) of Bitcoin for the time period. Figure \ref{hist_normal_returns} is the histogram and the normal quantile-quantile (q-q) plot of the return series for the same time period. The residuals plot is presented in figure \ref{returns_residuals_plot}. There is the presence of fatty tails and skewness in the residuals of the return series. This is confirmed from the Jacque-Bera and Anderson-Darling test statistics is table \ref{JB_AD_test_statistics}. Both tests rejected the normality at $5\%$ significance level. 

\begin{table}
	\caption{Bitcoin statistics, as at 01/09/2019}
	\centering
	\resizebox{\columnwidth}{!}{%
		\begin{tabular}{cccccccccc}		\hline 
			Cryptocurrency & \thead{Return on \\ investment (ROI)}  & \thead{Market \\ Capitalization} & \thead{All time \\ high} &  \thead{All time\\ low} \\ 
			\hline 
			Bitcoin &  7,005.34\%  & \$172,355,391, 698 \textsuperscript{} & \$20,089.00  & \$65.53  \\
			\midrule[.5pt]
			\multicolumn{4}{l}{\textsuperscript{}\footnotesize{Accessed at \url{https://coinmarketcap.com/} on 01/09/2019}}
			\\
		\end{tabular} 
	}
	\label{bitcoin_info}
\end{table}

\begin{table}[H]
	\caption{Descriptive statistics of the three major cryptocurrency by market capitalization}	\resizebox{\columnwidth}{!}{%
		\centering
		\begin{tabular}{cccccccccc}		\hline 
			Cryptocurrency & Maximum   &  Minimum  & Mean  & Standard deviation &  Skewness & Kurtosis \\ 
			\hline 
			Bitcoin (USD) & -0.8488 & 1.4744 & 0.0012 & 0.0600  & 216.7471 & 6.4820
			\\
			\hline 
		\end{tabular} 
	}
	\label{descriptive_statistics}
\end{table}

\begin{table}
	\caption{Jacque-Bera and Anderson-Darling test of normality for the residuals of the returns series}	
	\centering
	\begin{tabular}{cccc}		\hline 
		& Jacque-Bera test  & Anderson-Darling test  \\ 
		\hline 
		P-value & $< 2.2e-16$ & $< 2.2e-16$
		\\
		\hline 
	\end{tabular} 
	\label{JB_AD_test_statistics}
\end{table}

\noindent
The Augmented Dickey Fuller (ADF) test \cite{dickey1979distribution} is used to test for stationarity. From table \ref{arch_stationary_effects}, the null hypothesis of statonarity is rejected at 5\% $\alpha$-level of significance. Hence, there is no need to difference the return series. The Ljung-box and LM test are presented in table \ref{arch_stationary_effects}. From the Ljung box test, the null hypothesis of 'no autocorrelation' in the squared residuals is rejected at 5\% significance level. That is, there is dependency in the squared returns series of Bitcoin.  Using the Lagrange multiplier (LM) test, the null hypothesis of 'no ARCH effects' is rejected at 5\% significance level. From the Ljung box and LM test, it can be concluded that the volatility ARCH effect is very much present in the return series. Hence, the GARCH models to model the returns series data. 

\begin{table}[H]
	\caption{Test of Auto Regressive Conditional Heteroscedasticity (ARCH) effect }	
	\centering
	\begin{tabular}{cccc}		\hline 
		& Ljung box test  & LM test & ADF test \\ 
		\hline 
		P-value & $7.835e-05$  & $< 2.2e-16$ & 0.01
		\\
		\hline 
	\end{tabular} 
	\label{arch_stationary_effects}
\end{table}

\subsection{Estimated Volatility}
\begin{figure}[H]
	\centering
	\subfloat[sGARCH-Student t]{%
		\label{sGARCH_t_qq}%
		\includegraphics[height=05cm,width=4cm]{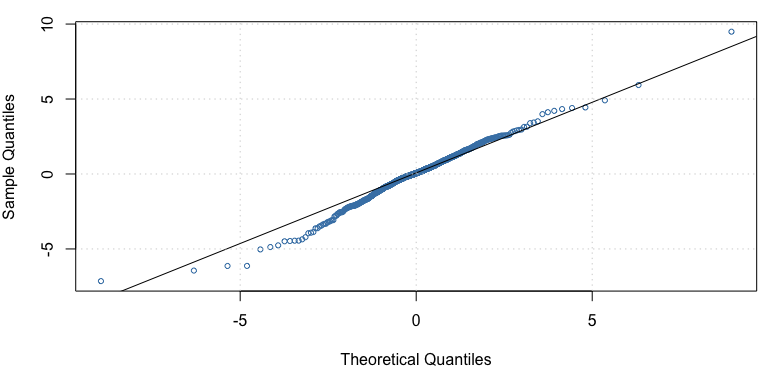}}%
	\qquad
	\subfloat[iGARCH-Student t]{%
		\label{iGARCH_t_qq}%
		\includegraphics[height=05cm,width=4cm]{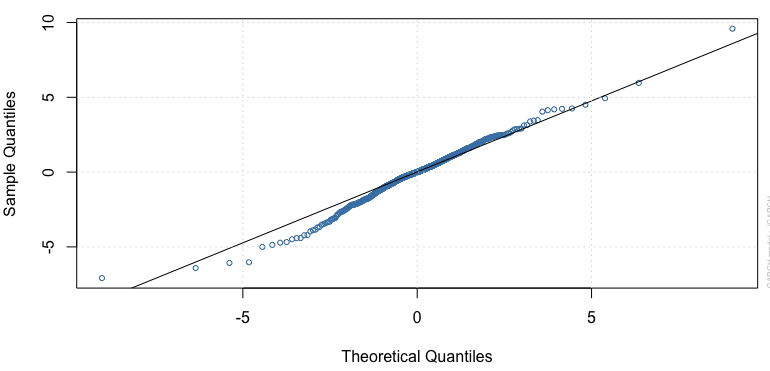}}%
	\qquad
	\subfloat[tGARCH-Student t]{%
		\label{tGARCH_t_qq}%
		\includegraphics[height=05cm,width=4cm]{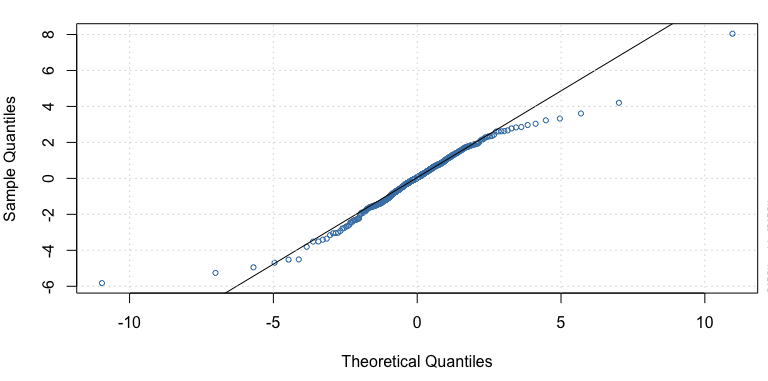}}%
	\caption{QQ plot of sGARCH, iGARCH, and tGARCH using the Student t-distribution (Student t)}
\end{figure}

\begin{figure}[H]
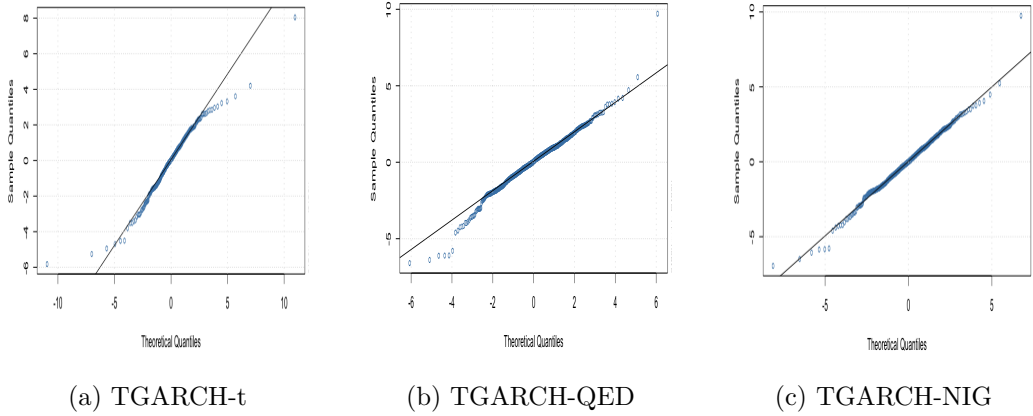

	\centering
	\subfloat[TGARCH-t]{%
		\label{tGARCH_t}%
		\includegraphics[height=05cm,width=4cm]{tGARCH_t_qq}}%
	\qquad
	\subfloat[TGARCH-QED]{%
		\label{tGARCH_qed}%
		\includegraphics[height=05cm,width=4cm]{tGARCH_qed_qq}}%
	\qquad
	\subfloat[TGARCH-NIG]{%
		\label{tGARCH_nig}%
		\includegraphics[height=05cm,width=4cm]{tGARCH_nig_qq}}%
	\caption{QQ plot using of the best distribution from sGARCH, iGARCH, and TGARCH}
	\end{figure}

\begin{figure}[H]
	\centering
\subfloat[Conditional volatility]{%
	\label{conditional_volatility}%
	\includegraphics[height=05cm,width=6cm]{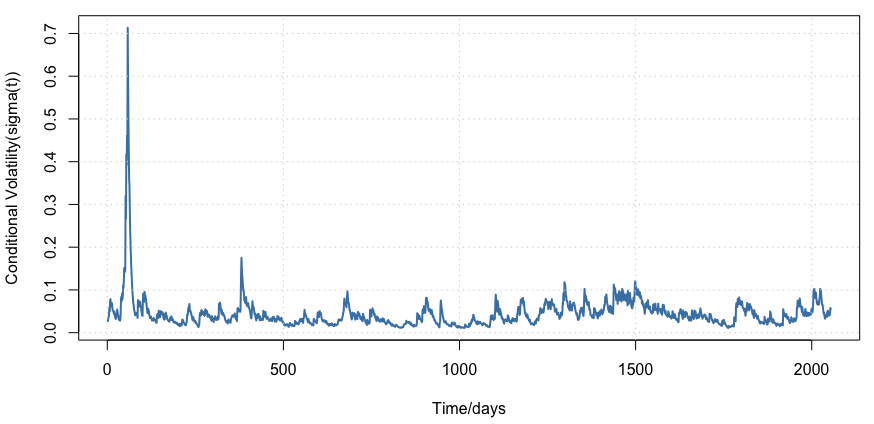}}%
\qquad
\subfloat[Density]{%
	\label{density}%
	\includegraphics[height=05cm,width=6cm]{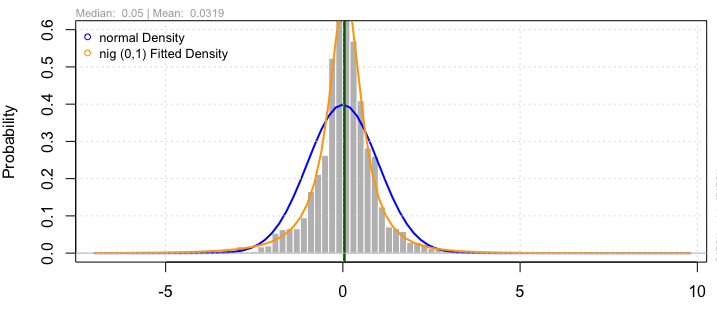}}%
\caption{Conditional volatility and density of  tGARCH(1,1)-NIG}	
\end{figure}

\noindent
Table \ref{student_t_tabular} shows the maximum likelihood estimate (MLE) results of sGARCH(1,1), iGARCH(1,1), and tGARCH(1,1) models for Bitcoin returns using student t-distribution. From the table, the log-likelihood value $(4181.104)$ is maximum for tGARCH(1,1) model. The values of the two information criterion (AIC=$-4.0663$, BIC=$-4.0526$) of tGARCH(1,1) are minimum as compared to sGARCH(1,1) and iGARCH(1,1). The visual QQ plot in Figure \ref{tGARCH_t_qq} is consistent with the AIC, BIC, and log likelihood values of the tGARCH(1,1). These results indicate that tGARCH(1,1) is the optimal model to describe the volatility of the return series of Bitcoin using the student t-distribution.  Table \ref{ged_tabular} presents MLE results of sGARCH(1,1), iGARCH(1,1), and tGARCH(1,1) models for Bitcoin returns using the generalized error distribution. Compared to the other models, the log-likelihood value is maximised under the tGARCH(1,1). The AIC and BIC values of tGARCH(1,1) is the minimal as compared to the other two models. Hence, the tGARCH(1,1) model is the best model relative to the sGARCH(1,1) and iGARCH(1,1). From Table \ref{nig_tabular}, the tGARCH(1,1) recorded the maximum log likelihood value (4196.681). The AIC and BIC values (-4.0805 and -4.0641 respectively) were also the lowest as compared to sGARCH(1,1) and iGARCH(1,1). This indicates that tGARCH(1,1) is the best model for the volatility of the return series of Bitcoin using the Normal Inverse Gaussian distribution.  
\\
From the three selected models (tGARCH(1,1) using student t-distribution, tGARCH(1,1) using generalized error distribution, and tGARCH(1,1) using normal inverse Gaussian distribution), the tGARCH(1,1) using Normal Inverse Gaussian distribution is the most optimal model for the volatility of Bitcoin return series. This is confirmed from the maximum log likelihood value and the minimum AIC and BIC values. This is clear from the visual plot in figure \ref{tGARCH_nig}. It is therefore evident that optimal model in terms of the information criterion, log likelihood, and QQ plot is the tGARCH(1,1) using the Normal Inverse Gaussian distribution. 
\\
The volatility estimates obtained from tGARCH(1,1)-NIG model is displayed in Figure \ref{conditional_volatility}. It is evident that volatility moves through time. The density of tGARCH(1,1)-NIG is shown in Figure \ref{density}. Clearly, the NIG distribution was able to capture the fat tails and skewness in the distribution. This confirms the reliability and efficiency in using the NIG distribution for modelling the volatility of Bitcoin return series.

\begin{table}[H]
\caption{MLE results of sGARCH(1,1), iGARCH(1,1), and tGARCH(1,1) models for Bitcoin returns with conditionally t-distributed errors}
	\renewcommand{\arraystretch}{1.5}
	\centering	
	\resizebox{\columnwidth}{!}{%
		\begin{tabular}{>{\bfseries}c*{17}{c}} 
			\toprule
	\multirow{2}{*}{\bfseries Model} &
			\multicolumn{5}{c}{} \\
\cmidrule(lr){3-7} \cmidrule(lr){8-11}
&& {$\hat{\omega}$} & {$\hat{\alpha}$} &  {$\hat{\beta}$} & {$1/v$} & {$\hat{\lambda}$} && {AIC} & {BIC} & {LogLikelihood}
\\ 
\cmidrule(lr){3-7} \cmidrule(lr){8-11}
sGARCH(1,1) && 0.000036 & 0.192227 & 0.806773 & 3.200514 & --- && -4.0367 & -4.0258 & 4149.72     
\\
&& (0.000012) & (0.023060) & (0.023828) & (0.169435) & --- && &  &      
			\\
iGARCH(1,1) && 0.000035 & 0.192695 & 0.807305 & 3.194742 & --- && -4.0379 & -4.0297 & 4149.946    
\\
&& (0.000010) & (0.021763) & --- & (0.135971) & --- && &  &        
			\\
tGARCH(1,1) && 0.001019 & 0.291106 & 0.835797 & 2.472141 & 0.007515 && \bf{-4.0663} & \bf{-4.0526} & \bf{4181.104}             
			\\
&& (0.000353) & (0.048010) & (0.018121) & (0.157392) & (0.059760) && &  & 
\\
			\bottomrule
\multicolumn{4}{l}{\textsuperscript{}\footnotesize{Note: Standard errors of estimates are reported in parentheses.}}
\\
\multicolumn{4}{l}{\textsuperscript{}\footnotesize{Note 2: {$\hat{\omega}$:The reaction of conditional variance}, {$\hat{\alpha}$:ARCH effect}, {$\hat{\beta}$:GARCH effect}, {$\hat{\lambda}$:Leverage effect}}}
		\end{tabular}
	}
	\label{student_t_tabular}
\end{table}

\begin{table}[H]
	\caption{MLE results of sGARCH(1,1), iGARCH(1,1), and tGARCH(1,1) models for Bitcoin returns with conditionally generalized error distribution errors}
	\renewcommand{\arraystretch}{1.5}
	\centering	
	\resizebox{\columnwidth}{!}{%
		\begin{tabular}{>{\bfseries}c*{17}{c}} 
			\toprule
			\multirow{2}{*}{\bfseries Model} &
			\multicolumn{4}{c}{\bfseries  Estimated parameters} &
			\multicolumn{5}{c}{} \\
			\cmidrule(lr){3-7} \cmidrule(lr){8-11}
&& {$\hat{\omega}$} & {$\hat{\alpha}$} &  {$\hat{\beta}$} & {$1/v$} & {$\hat{\lambda}$} && {AIC} & {BIC} & {LogLikelihood}
			\\ 
				\cmidrule(lr){3-7} \cmidrule(lr){8-11}
sGARCH(1,1) && 0.000039 & 0.192872 & 0.806128 & 0.860227 & --- && -4.0495 & -4.0386 & 4162.87      
			\\
&& (0.000011) & (0.025780) & (0.022277) & (0.030944) &  --- && &  &      
			\\
iGARCH(1,1) && 0.000039 & 0.193739 & 0.806261 & 0.859610 & --- && -4.0506 & -4.0424 & 4162.95      
			\\
&& (0.000010) & (0.022141) & --- & (0.029442)& --- && &  &        
			\\
tGARCH(1,1) && 0.001298 & 0.212642 & 0.826352 & 0.851479 & 0.055244 && \bf{-4.0632} & \bf{-4.0495} & \bf{4177.926}              
			\\
&& (0.000337) & (0.025368) & (0.021071) & (0.031225) & (0.061816) && &  & 
			\\
			\bottomrule
\multicolumn{4}{l}{\textsuperscript{}\footnotesize{Note1: Standard errors of estimates are reported in parentheses.}}
		\end{tabular}
	}
	\label{ged_tabular}
\end{table}

\begin{table}[H]
	\caption{MLE results of sGARCH(1,1), iGARCH(1,1), and tGARCH(1,1) models for Bitcoin returns with conditionally normal inverse gaussian distribution errors}
	\renewcommand{\arraystretch}{1.5}
	\centering	
	\resizebox{\columnwidth}{!}{%
		\begin{tabular}{>{\bfseries}c*{17}{c}} 
			\toprule
			\multirow{2}{*}{\bfseries Model} &
			\multicolumn{5}{c}{\bfseries  Estimated parameters} &
			\multicolumn{5}{c}{} \\
			\cmidrule(lr){3-8} \cmidrule(lr){9-12}
&& {$\hat{\omega}$} & {$\hat{\alpha}$} &  {$\hat{\beta}$} & {$1/v$} &  
			{$\hat{\kappa}$} & {$\hat{\lambda}$} && {AIC} & {BIC} & {LogLikelihood} 
			\\ 
			\cmidrule(lr){3-8} \cmidrule(lr){9-12}
			sGARCH(1,1) && 0.00004 & 0.19748 & 0.80152 & 0.39161 & -0.13314 & --- && -4.0599 & -4.0462 & 4174.485     
			\\
			&& (0.000012) & (0.024326) & (0.022933) & (0.046759) & (0.032272) & --- &&  &      &
			\\
			iGARCH(1,1) && 0.000039 & 0.198112 & 0.801888 & 0.390387 & -0.133277 & --- && -4.061 & -4.050  & 4174.653      
			\\
			&& (0.000010) & (0.022055) & --- & (0.042123) & (0.032213) & --- &&  &        &
			\\
			tGARCH(1,1) && 0.001048 & 0.233978 & 0.828983 & 0.319214 & -0.142017 & -0.005427 && \bf{-4.0805} & \bf{-4.0641} & \bf{4196.681}        
			\\
			&& (0.000302) & (0.027732) & (0.019678) & (0.042593) & (0.033599) & (0.061556) && &  & 
			\\
			\bottomrule
\multicolumn{4}{l}{\textsuperscript{}\footnotesize{Note: Standard errors of estimates are reported in parentheses.}}
		\end{tabular}
	}
	\label{nig_tabular}
\end{table}

\begin{figure}[H]
	\centering
	\subfloat[sGARCH-GED]{%
		\label{sGARCH_qed_qq}%
		\includegraphics[height=05cm,width=4cm]{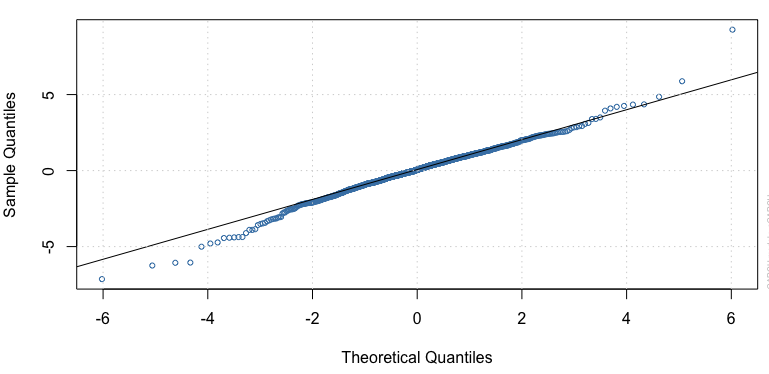}}%
	\qquad
	\subfloat[iGARCH-GED]{%
		\label{iGARCH_ged_qq}%
		\includegraphics[height=05cm,width=4cm]{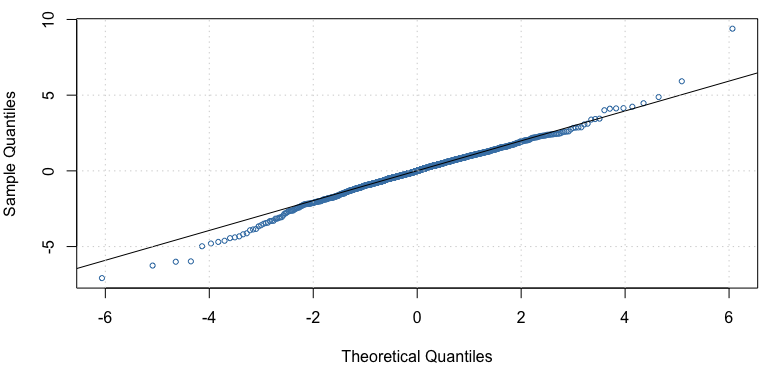}}%
	\qquad
	\subfloat[tGARCH-GED]{%
		\label{tGARCH_qed_qq}%
		\includegraphics[height=05cm,width=4cm]{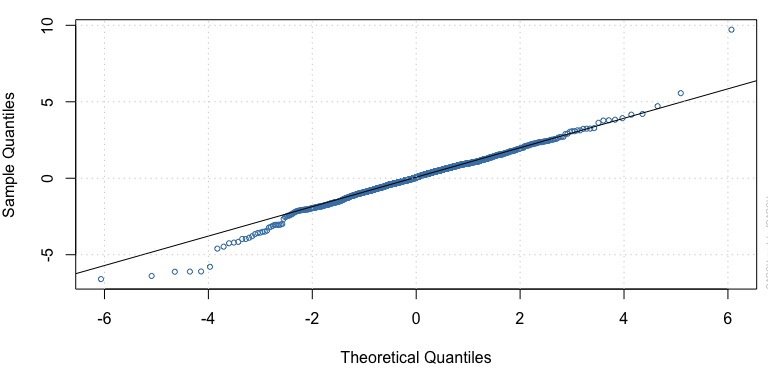}}%
	\caption{QQ plot of the sGARCH, iGARCH, and tGARCH using the Generalized Error Distribution (GED)}
\end{figure}

\begin{figure}[H]
	\centering
	\subfloat[sGARCH-NIG]{%
		\label{sGARCH_nig_qq}%
		\includegraphics[height=05cm,width=4cm]{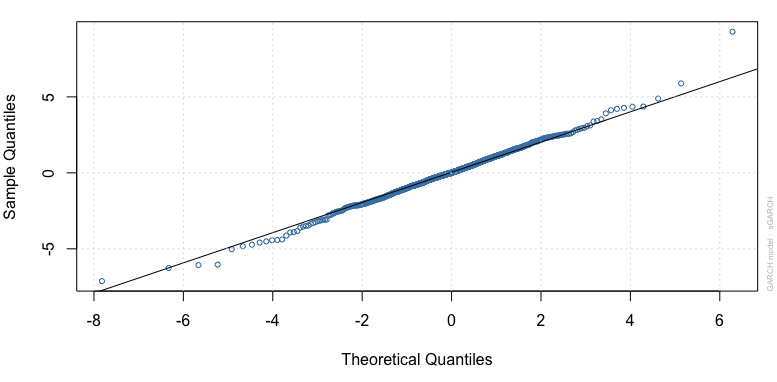}}%
	\qquad
	\subfloat[iGARCH-NIG]{%
		\label{iGARCH_nig_qq}%
		\includegraphics[height=05cm,width=4cm]{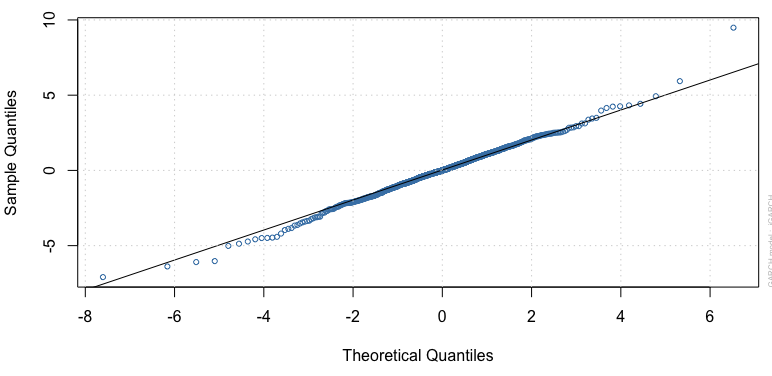}}%
	\qquad
	\subfloat[TGARCH-NIG]{%
		\label{tGARCH_nig_qq}%
		\includegraphics[height=05cm,width=4cm]{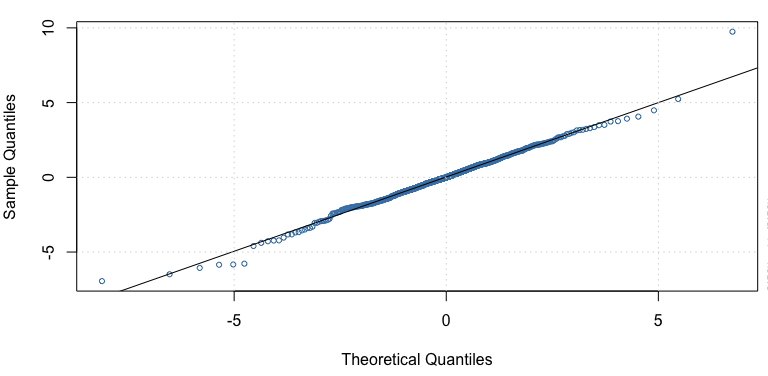}}%
	\caption{QQ plot of sGARCH, iGARCH, and tGARCH using the Normal inverse Gaussian (NIG) distribution }
\end{figure}

\section{Conclusion}
\noindent
This paper studied the volatility of daily return series of Bitcoin from 01/01/2014 to 16/08/2019. The results of the statistical properties revealed that just like other financial time series data, the return series of Bitcoin are leptokurtic. Different GARCH type models (sGARCH, iGARCH, tGARCH) were compared and the tGARCH model was identified to be the most appropriate model for estimating the time-varying volatility in Bitcoin return series. To account for the skewness and fat tails, all the GARCH type models were compared based on Student-t, Generalized Error Distribution (GED), and Normal Inverse Gaussian (NIG) distribution. The NIG distribution performed better in capturing the fat tail and skewness in the return series distribution. Hence, the tGARCH-NIG model was the optimal model for modeling and estimating the volatility in the return series of Bitcoin.

\section*{Conflict of Interest}
\noindent
The author declare that there are no conflicts of interest regarding the publication of this paper.

\section*{Data Availability}
\noindent
Data for this work are available from the author upon request.


\bibliographystyle{model1-num-names}
\bibliography{sample.bib}

\end{document}